\documentclass[10pt,prb,aps,twocolumn,superscriptaddress,floatfix,notitlepage,nofootinbib,longbibliography]{revtex4-1}
\usepackage[latin1]{inputenc}
\usepackage{mathrsfs}
\usepackage{amsmath,dsfont}
\usepackage{amsfonts,natbib}
\usepackage{amssymb}
\usepackage{bm}
\usepackage[pdftex]{graphicx}
\usepackage{comment}

\usepackage{marginnote}

\usepackage{amssymb,color}

\usepackage[colorlinks=true,allcolors=blue]{hyperref}

\begin{document}
	\title{New Kind of Echo from Quantum Black Holes} 
	\author{Sreenath K. Manikandan}
	\email{sreenath.k.manikandan@su.se}
	\affiliation{Nordita,
KTH Royal Institute of Technology and Stockholm University,
Hannes Alfv\'{e}ns v\"{a}g 12, SE-106 91 Stockholm, Sweden}
	\author{Karthik Rajeev}
	\email{karthikrajeev.kr@gmail.com}
	\affiliation{School of Physical Sciences, Indian Association for the Cultivation of Science, Kolkata-700032, India}
	\affiliation{Department of Physics, Indian Institute of Technology Bombay, Mumbai, 400076, India}
	
	\date{\today}
	\begin{abstract}
We propose that a quantum black hole can produce a new kind of  late-time gravitational echoes, facilitated by a near-horizon process analogous to Andreev reflection in condensed matter systems. In comparison to the traditional echo scenarios where the near-horizon region is treated as an ordinary reflector, we argue that, consequent to near-horizon gravitational scattering, this region is better described by an Andreev reflector. Such interactions lead to a novel contribution to gravitational echoes with a characteristic phase difference, an effect which is analogous to how Andreev reflections lead to propagating particle-like and hole-like components with a relative phase in certain condensed matter scenarios. Moreover, this novel contribution to the echo signal encodes information about the `near-horizon quantum state', hence offering a possible new window to probe the quantum nature of black holes. 
	\end{abstract}
	\maketitle
\section{Introduction} 
Black holes, for several reasons, are one of the most peculiar objects in nature. On the one hand, as mathematical constructions in general relativity (GR), they are remarkably simple in their rendition~\cite{wheeler1971introducing}, while on the other, they seem to possess thermodynamic properties~\cite{Bekenstein:1972tm,Bardeen:1973gs,Hawking:1974rv} which are usually ascribed to objects that have a microscopic structure. Over the years, the studies on black holes have bestowed us with insights into several branches of theoretical physics and mathematics. Moreover, with several recent~\cite{LIGOScientific:2018mvr,LIGOScientific:2020ibl,LIGOScientific:2021djp} and upcoming~\cite{Bailes:2021tot} gravitational-wave observations directly concerning black holes, it is expected that they may prove to be useful in verifying many important theoretical predictions in the field. A deeper understanding of black hole physics is also particularly crucial to the efforts towards a consistent theory of quantum gravity. In this regard, despite years of investigations, there is hardly any dispute that much is yet to be deciphered about black holes with quantum characteristics.

The most notorious puzzle in the quantum physics of black holes is the black hole quantum information loss problem~\cite{Hawking:1976ra,Mathur:2009hf,Almheiri:2012rt}; it concerns the complete recovery of the initial quantum state that collapsed to form a black hole, from the radiation which is left behind and available to an asymptotic future observer. Several approaches have been suggested over the years to resolve this issue (see~\cite{Harlow:2014yka,Chakraborty:2017pmn,Perez:2017cmj,Raju:2020smc,RevModPhys.93.035002} for a recent review). 
Insights from condensed matter platforms have also resulted in intriguing quantum analogies which suggest that the final quantum state of black holes as perceived by an external observer could be a superfluid quantum condensate and, hence, leading to a viable alternative paradigm to understand the black hole evaporation process ~\cite{Manikandan:2018urq,Manikandan:2017zhw,Manikandan:2020kdx}.
  
\begin{figure*}[t]
\centering
\includegraphics[width=\linewidth]{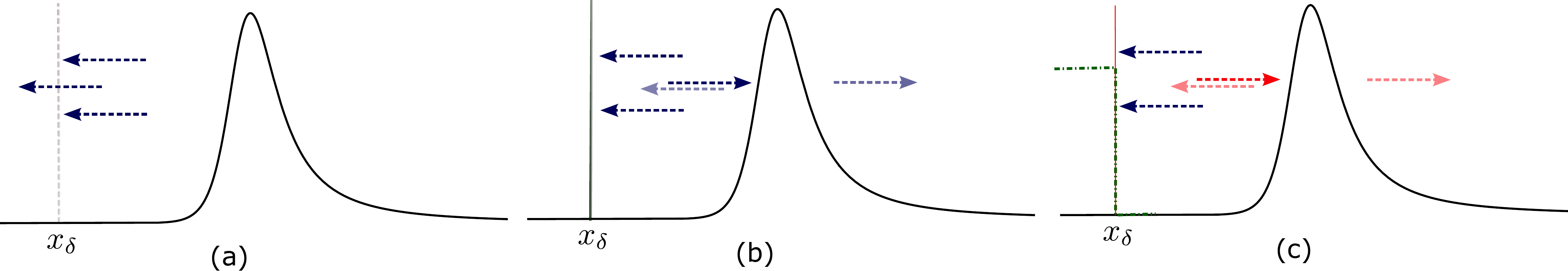}
\caption{(a) Black hole horizon as a perfect absorber, where all the modes propagating to the left of the potential barrier (with $l=2,~s=2$ in this example) are absorbed. (b) Echoes are created when the boundary condition in the near horizon region $x=x_{\delta}$ is replaced by a reflector. The reflectivity depends on the specifics of the modified gravity theory under consideration, for instance the proposal in Refs.~\cite{Bekenstein:2020mmb,Bekenstein:1995ju} suggests that the horizon area is quantized, such that the reflectivity is unity except for at the discrete frequencies given in Eq.~\eqref{spec}, where one expects absorption to occur. (c) Our proposal, where quantum interactions near the horizon facilitates  mode conversions, leading to both particle-like and hole-like components for the echo.}
\label{echo}
\end{figure*} 

Further inputs to the notion that a black hole could be a condensate in a consistent quantum theory of gravity also comes from several independent considerations, see for instance~\cite{Dvali:2012rt,Dvali:2012gb,Dvali:2012en,Dvali:2011aa,Jacobson:1996zs}. Notwithstanding the elegance of such proposals and their conceptual implications, any consistent quantum description of black holes should yield measurable predictions. With the growing capabilities of gravitational-wave and other observations, there is increasing confidence that substantial inputs for the formulation of a quantum theory of gravity maybe available in the decades to come~\cite{Amelino-Camelia:2009imt,Pikovski:2011zk,PhysRevD.99.124053,Bekenstein:2013ih,Bekenstein:2012yy,Bosso:2016ycv}. In view of this, through this work, we highlight the observational aspects of certain proposals that suggest a black hole could be a condensate of interacting microscopic degrees of freedom. In particular, since the quantum state of a condensate is characterized by very few parameters, as for instance the number density and macroscopic quantum phase of the condensate, such proposals are likely to have effective descriptions with minimal number of parameters. Moreover, if available or upcoming observational facilities can probe the macroscopic quantum features of the condensate, one might gain useful insights about the quantum nature of black hole horizons. Here, we propose a scenario which precisely addresses this possibility, using the framework of the so-called gravitational echoes from a black hole spacetime~\cite{PhysRevLett.116.171101,Cardoso:2016oxy,Cardoso:2017cqb,Cardoso:2019apo,Wang:2019rcf}.

The possibility of creation of multiple echoes of gravitational signals from black holes and exotic compact binaries has caught significant attention in recent times~\cite{PhysRevLett.116.171101,Cardoso:2016oxy,Cardoso:2017cqb,Cardoso:2019apo,Wang:2019rcf,Dong:2020odp}.  These echoes arise as a result of ordinary reflective boundary conditions imposed at the horizon. Multiple reflections of gravitational waves at the horizon and the photon sphere potential barrier results in a signal pattern reminiscent of the echoes formed as a result of multiple reflections of sound waves. Imposition of non-standard reflective boundary conditions at horizon could be motivated by several proposed modifications of the black hole spacetime that arise, for instance, from quantum gravity considerations. For example, recently authors of Ref.~\cite{Cardoso:2019apo} delineated an approach to verify the proposal in Refs.~\cite{Bekenstein:2020mmb,Bekenstein:1995ju} that a quantum black hole has quantized horizon area spectrum, using gravitational wave observations. We review their approach in Sec.~\ref{gechoes}.

In this article, we analyze the consequence of applying a different kind of boundary condition at the event horizon where the near horizon region is treated as an Andreev reflector~\cite{Manikandan:2017zhw,Manikandan:2018urq,Manikandan:2020kdx,Jacobson:1996zs,osti_4071988}. While this consideration is primarily motivated by proposals that treat an evaporating black hole as a leaking superfluid quantum condensate, such modifications may also be understood as emerging from the gravitational self interaction of test fields in the background of an evaporating black hole. In particular, when the test field is a tensor mode of perturbation, our analysis suggests a fundamentally new kind of gravitational wave echo that can be detected. Such modified boundary conditions may also be of relevance to exotic compact objects other than black holes such as neutron stars~\cite{Pani:2018flj,Urbano:2018nrs,Maggio:2019zyv}. See Fig.~\ref{echo} where we compare different echo frameworks in comparison to our proposal.

The article is organized as follows. In Sec.~\ref{gechoes}, we review the general notion of echoes from a black hole in gravitational quantum physics. In Sec.~\ref{sect_condensate}, we summarize arguments from the gravitational side which motivates our treatment of the near-horizon region as a mode-converting (Andreev reflecting) condensate.  In Sec.~\ref{andreevbosons}, we review the Andreev reflection mechanism for a condensate of superfluid bosons. In Sec.~\ref{newkind}, we discuss how Andreev reflection can provide a novel contribution to the echo from a black hole spacetime. In Sec.~\ref{discuss}, we conclude by discussing the implications of our prediction for near-future gravitational observations.

\section{Gravitational echoes\label{gechoes}}
The relevant region of spacetime that we are focusing on is described by the following standard Schwarzschild metric:
\begin{align}
    ds^{2}=-\left(1-{\frac {2GM}{r}}\right)\,dt^{2}+\left(1-{\frac {2GM}{r}}\right)^{-1}\,dr^{2}+r^{2}d\Omega ^{2}.
\end{align}
The master equation describing the dynamics of the perturbations of massless fields in the background of the Schwarzschild metric reads~\cite{Berti:2009kk,Regge:1957td,Thorne:1980ru,Fiziev:2005ki,Cardoso:2019apo}:
\begin{align}\label{regge_wheeler}
\left[\partial^2_t-\partial^2_x+V_l(x)\right]\psi(x,t)=\mathcal{S}(x,t),
\end{align}
where $\mathcal{S}(x,t)$ denotes a source term and we have also introduced the tortoise coordinate $x$ via $x=r+2M\log(r/2M-1)$. The effective potential $V_l$ takes the form (in the original Schwarzschild radial coordinate $r$)~\cite{Berti:2009kk,Regge:1957td,Cardoso:2019apo,Fiziev:2005ki,Thorne:1980ru}:
\begin{align}
    V_l(r)=\left(1-\frac{2M}{r}\right)\left[\frac{l(l+1)}{r^2}+\frac{(1-s^2)2M}{r^3}\right].
\end{align}
Here $s=0,1,2$ respectively corresponds to scalar, electromagnetic and gravitational perturbations. 

To resolve the dynamics of test fields described by Eq.~\eqref{regge_wheeler}, one can Fourier analyse the components of the test field $\psi(x,t)$. This results in the following form of Eq.~\eqref{regge_wheeler} in the Fourier space~\cite{Cardoso:2019apo},
\begin{align}\label{regge_wheelerw}
\left[\omega^{2}-\partial^2_x+V_l(x)\right]\tilde{\psi}(x,\omega)=\tilde{\mathcal{S}}(x,\omega).
\end{align}
A requirement owing to causality is that the test fields $\psi(x,t)$ in the time domain obey the Sommerfeld boundary conditions $\partial_{t}\psi+\partial_{x}\psi=0$ as $x\rightarrow\infty$. This translates to the effect that the Fourier components of test fields  $\tilde{\psi}(x,\omega)$, as $x\rightarrow\infty$, behave as $\tilde{\psi}(x,\omega)\propto e^{i\omega x}$~\cite{Cardoso:2019apo}. 

The near horizon region is completely absorbing for a classical black hole, and therefore one  traditionally assumes a completely ingoing boundary condition for test fields at the event horizon. 
It has been suggested that quantum mechanical corrections to the physics of the event horizon may challenge this notion. A well-known example to such a modification is the model pioneered in Refs.~\cite{Bekenstein:2020mmb,Bekenstein:1995ju}, which suggests that the Horizon area $A$ is quantized,
\begin{equation}
    A=\alpha l_{p}^{2} n,
\end{equation}
where $l_{p}=\sqrt{\hbar G/c^2}$ is the Planck length, $\alpha$ is a dimensionless coefficient (there is some indication that $1<\alpha<30$, see Ref.~\cite{Cardoso:2019apo}), and $n$ is an integer. Such a quantization of the area spectrum implies that the black hole area (and entropy thereof) can only change in discrete units. Subsequently, the frequency of test fields absorbed or emitted by a Schwarzschild black hole also has a discrete spectrum, given by~\cite{Bekenstein:2020mmb,Bekenstein:1995ju,Cardoso:2019apo},
\begin{equation}
    \omega_{n} = \frac{\alpha c\delta n}{16\pi r_{h}},\label{spec}
\end{equation}
where $\delta n$ indicates the change in the area quantum,  $c$ is the speed of light, and $r_{h}$ is the Schwarzschild radius.   A discussion of more general discrete spectrum of black holes can be found in Ref.~\cite{Lochan:2015bha}.

  Such a quantum gravity modification for the near-horizon region implies that the Fourier components of test fields $\tilde{\psi}(x,\omega)$ satisfy a different boundary condition in the near-horizon region, given by~\cite{Cardoso:2019apo},
\begin{equation}
    \tilde{\psi}(x,\omega)\propto e^{i\omega x}+R(\omega)e^{-i\omega x}\quad;\quad x\sim x_{\delta}, \label{echoOld}
\end{equation}
where $R(\omega)$ is the modified reflectivity of the near-horizon region and $x_{\delta}$ is the position in tortoise coordinates corresponding to the Schwarzschild radial coordinate $r=r_{h}+\delta$, satisfying $\delta/r_{h}\ll 1$. For the proposal in Refs.~\cite{Bekenstein:2020mmb,Bekenstein:1995ju}, the reflectivity is expected to go to unity except for the discrete frequencies 
given in Eq.~\eqref{spec} where absorption lines are expected~\cite{Cardoso:2019apo}. 

A consequence of modified reflectivity $R(\omega)$ of the near-horizon region is the creation of gravitational echoes of test fields; the potential barrier $V_l(r)$ partially reflects and partially transmits test fields that are emanated or reflected from the near-horizon region, leading to an echo-like signal. As multiple reflections and transmissions occur, it is expected that the potential barrier $V_l(r)$ will have a filtering effect on the measured signal. For the proposal in Refs.~~\cite{Bekenstein:2020mmb,Bekenstein:1995ju}, the measured signal at later times will have sharp absorption peaks at the frequencies $\omega=\omega_{n}$, given in Eq.~\eqref{spec}. As the test field can also be a gravitational perturbation, the prediction in Ref.~\cite{Cardoso:2019apo} is that the above filtering effect may be observed in gravitational wave detectors, and therefore could serve as a potential way to test the black hole area quantization proposed in Refs.~\cite{Bekenstein:2020mmb,Bekenstein:1995ju}. 

 A recent work has in fact looked into the possibility of constraining the parameter $\alpha$ based on presently available gravitational wave observations~\cite{Laghi:2020rgl}. The analysis therein suggests that the information available from gravitational wave observations until October $1$,
2019 is not sufficient enough to fully support or disregard the proposal in Ref.~\cite{Cardoso:2019apo}.   It is also worth mentioning that the black hole area quantization also has interesting consequences to the inspiral phase, as has been discussed in Ref.~\cite{Datta:2021row}.

Before we conclude this section, we would also like to briefly comment on the source term in Eq.~\eqref{regge_wheeler}. In the first order in perturbation, a non-zero value of the source term $\mathcal{S}(x,t)$ signifies the presence of charges (like, for instance, the electromagnetic charge or mass) in the exterior of the black hole region. In higher order in perturbation, however, the component of the stress-energy that corresponds to self-interaction can also contribute to the source term. In the specific case of gravitational perturbations, such a contribution can arise from the gravitational-wave stress-energy and leads to, for instance, the well-known non-linear memory effect~\cite{Christodoulou:1991cr,Payne:1983rrr,Blanchet:1992br,Wiseman:1991ss}. Now imagine that the near horizon region, in fact, consists of a condensate of a particular matter field. In this case, we expect that the master equation for perturbation modes of that field contains a source term that corresponds to the contribution from the condensate. In the remainder of this article, we shall propose a toy model to study such a system.  
 
 \section{Near horizon interactions and the condensate picture}\label{sect_condensate}
\begin{figure}[t]
\centering
\includegraphics[scale=.5]{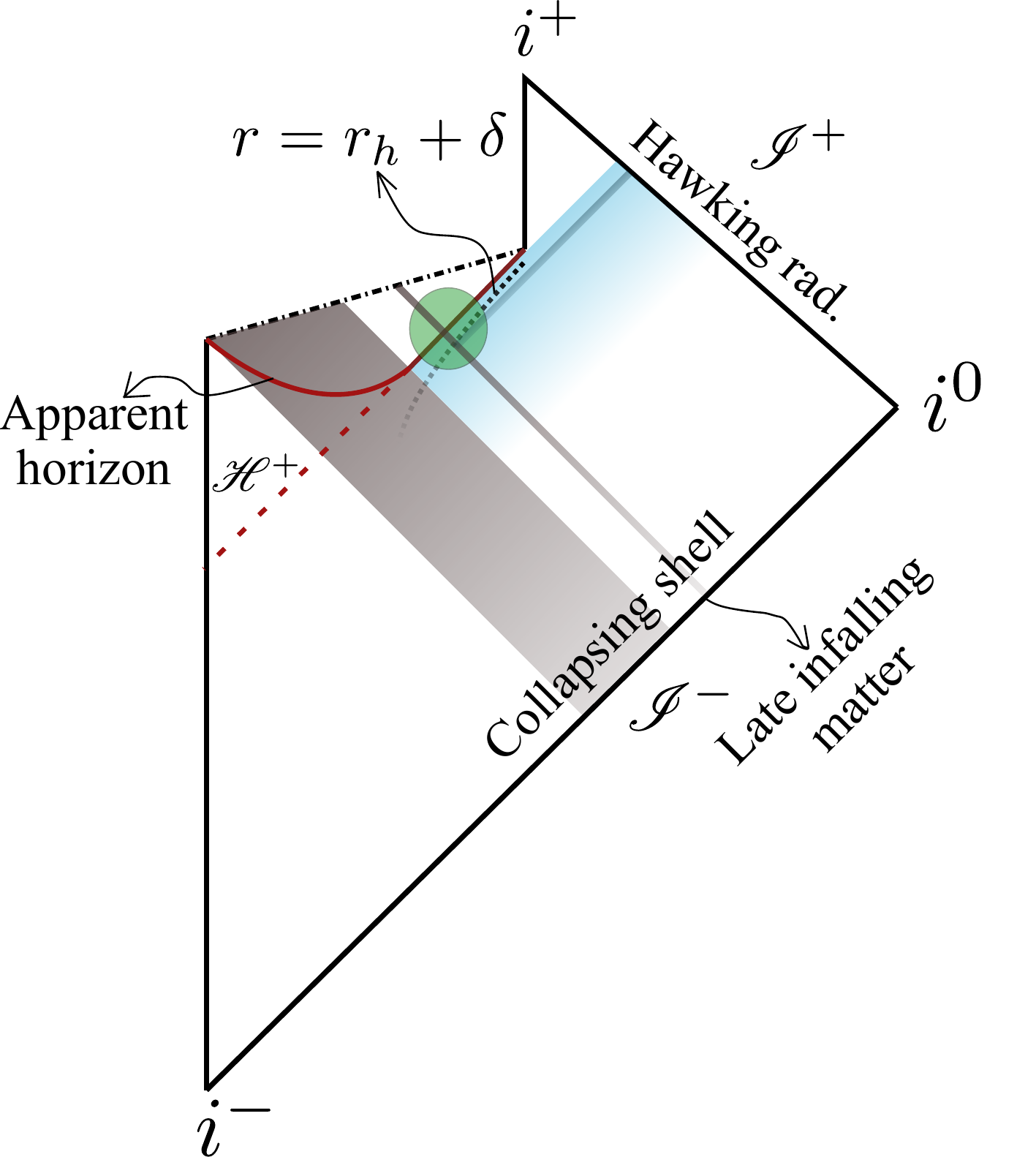}
\caption{The Penrose diagram for an evaporating black hole formed as a result of a spherical null-shell collapse. The dotted line at $r=r_{h}+\delta$, with $r_{h}$ being Schwarzschild radius models the interface of the strongly/weakly interacting regions of the Hawking radiation. Alternatively, such an interface may also be relevant in approaches that model the black hole as a graviton condensate, wherein the same acquires the interpretation of a super-fluid/normal-fluid regions (see text for more detail). The focus of this work is to model the scattering of a late in-falling mode at the near-horizon region inside the the green shaded circle.}
\label{penrose}
\end{figure}

We begin by considering a simple collapsing scenario, namely, the spherical collapse of a shell of massless scalar particles leading to the formation of a black hole. As is well known, the classical geometry describing this process can be obtained by stitching together patches of three exact solutions---(1) Minkowski spacetime (vacuum region inside the shell), (2) ingoing vaidya spacetime (non-vacuum region inside the shell) and (3) the Schwarzschild spacetime (exterior of the shell). However, as conveyed by Hawking's seminal semiclassical arguments, the black hole also evaporates by emitting near-thermal radiation and in due course exhaust all of it's mass (see Fig.~\ref{penrose}). We shall shortly focus on the in-fall of a massless scalar field mode (denoted by late infalling matter in Fig.~\ref{penrose}) into the black hole, long after the formation of event horizon and, at the same time, early enough to have not much of the black hole mass evaporated away. Hence, the spactime region of our interest lies somewhere inside the green circular region in Fig.~\ref{penrose}.

The fact that the state of radiation in the asymptotic future is non-vacuum is usually discerned from the non-trivial Bogoliubov transformations connecting the appropriate `in' and `out' modes, as was done in Hawking's seminal work~\cite{Hawking:1975vcx}. To calculate the corresponding Bogoliubov coefficients, one first traces the evolution of positive energy modes on $\mathscr{I}^{+}$ into the past, all the way to $\mathscr{I}^{-}$. This procedure leads to a relation between the positive energy modes on $\mathscr{I}^{+}$ and $\mathscr{I}^{-}$, from which one can derive the Bogoliubov coefficients and, hence, the particle spectrum. However, the tracing the evolution of the out-modes, in the conventional geometric optics approximation, ignores the gravitational interaction between the scalar field modes. It has been suggested that as a consequence of graviton mediated interactions, the Bogoliubov coefficients connecting the `in' and `out'-modes get modified, due to the non-trivial near-horizon scattering amplitude; it is also hoped that such a scenario leads to the resolution of the black hole quantum information loss problem~\cite{tHooft:1996rdg,tHooft:2015pce,Gaddam:2020mwe,Betzios:2016yaq,Betzios:2020xuj}.

The gravitational interaction may also significantly change the dynamics of the late in-falling matter that just crosses the horizon after passing through the outgoing Hawking radiation. For convenience, we may regard two distinct manners in which the gravitational interaction manifest in this scenario. Firstly, owing to the graviton mediated interaction, the vacuum appropriate to $\mathscr{I}^{-}$ will evolve into a state that is quite different from the one generated by a simple Bogoliubov mapping of the kind considered originally by Hawking in~\cite{Hawking:1974rv}. Secondly, the gravitational field of the in-falling matter may engrave, on the Hawking radiation, the information about the in-falling state. A rather rudimentary picture for such an interaction could be visualized in the following way: the spacetime gets slightly modified by the infalling matter, leading to a slight deviation in the trajectory of the outgoing quanta and vice versa. A more formal realization of this picture was considered in Ref.~\cite{Dray:1984ha} to effectively capture the near-horizon scattering of Aichelburg-Sexl gravitational shockwaves. 

We shall now present a simple model to study the near-horizon dynamics of a late in-falling mode, that incorporates the two above-mentioned effects of graviton mediated interactions. We start with the reasonable assumption that there is a critical distance beyond the horizon up to which the gravitational scattering between the scalar modes becomes relevant. In terms of the standard Schwarzschild radial coordinate $r$, with the event horizon at $r_h$, we shall assume that this critical region of significant gravitational scattering to be within $r=r_h+\delta$, demarcated by the dotted curve in Fig.~\ref{penrose}. Therefore, the outgoing Hawking radiation has two distinct parts: (1) in which the graviton exchange among the scalar field modes is significant ($r<r_h+\delta$) and (2) in which the scalar field modes can be safely assume to be freely propagating in the background black hole geometry. 

A brief detour to a problem in condensed matter is in order~\cite{Jacobson:1996zs,Manikandan:2017zhw,Manikandan:2018urq,Manikandan:2020kdx}. Recall that a superconductor-normal metal interface is characterized by a region of rapid decrease in the effective coupling of the phonon mediated interaction between electrons. One the superconductor side, the interaction gives rise to a gap, while the same is absent in the normal metal side. An analogous scenario occurs in a superfluid/normal fluid interface as well~\cite{2009PhRvL.102r0405Z}. In light of this, here we explore the possibility of modelling the suface at $r=r_h+\delta$, separating the outgoing Hawking radiation, as an interface between a condensate in the region $r<r_h+\delta$ and a coherent, almost free, distribution of particles in the region $r>r_h+\delta$. We also henceforth refer to the interactions within the aforementioned region as the ``horizon proximity effect", motivated by the analogy to the condensed matter setting~\cite{Manikandan:2017zhw,Manikandan:2018urq,Manikandan:2020kdx,Jacobson:1996zs}. 

Since we are considering the matter degree of freeedom to be a massless scalar field, it is reasonable to imagine the condensate (in $r<r_h+\delta$) as a superfluid and the outside Hawking radiation (in $r>r_h+\delta$) as the normal fluid. A remarkable consequence of this model is that the near horizon scattering of a late in-falling mode, with the outgoing Hawking radiation, can be modelled as the scattering problem near a superfluid/normal fluid interface. Note that our analysis suggests possible generalizations to electromagnetic and gravitational perturbations as well.
 The considerations that are to follow may also be applied to models that treat the black hole as a graviton condensate, for instance, along the lines of~\cite{Dvali:2011aa,Dvali:2012en}. 

In order to provide a better understanding of how assuming a quantum condensate description for the near-horizon region modifies the dynamics of test fields, we now look at the exact dynamics of quantum fluctuations of a condensate at a superfluid/normal fluid boundary. To this end, we shall closely follow the approach in~\cite{2009PhRvL.102r0405Z}.

\section{Bosonic analogue of Andreev reflection\label{andreevbosons}}
The dynamics of quantum fluctuations of the condensate in superfluid/normal fluid boundaries can be described by the Gross-Pitaevskii equation~\cite{gross1961structure,Pitaevskii1961}:
\begin{eqnarray}
    i\partial_t\Psi(x,t)&=&-\frac{1}{2}\partial^2_x\Psi(x,t)+V(x)\Psi(x,t)\nonumber\\&+&g(x)\left|\Psi(x,t)\right|^2\Psi(x,t),\label{GP}
\end{eqnarray}
which is a non-linear generalization of the Schr\"{o}dinger equation that accounts for inter-particle interaction, with $g(x)$ being the varying coupling strength. Now, let us consider a solution of the above equation that describes a small perturbation about an equilibrium wave function $\Psi_0$ describing a condensate. Such a solution takes the following form~\cite{2009PhRvL.102r0405Z}:
\begin{align}
    \Psi(x,t)=\Psi_0(x) e^{-i\mu t}+\delta\Psi.
\end{align}
The perturbation $\delta\Psi$, in turn, can be assumed to have the form~\cite{2009PhRvL.102r0405Z,Zapata:2011ze}:
\begin{align}
    \delta\Psi=e^{-i\mu t}\sum_{j}\left[u_{j}(x)e^{-i\omega_j t}-v^*_{j}(x)e^{i\omega_j t}\right],
\end{align}
with $u_j$ and $v_j$ satisfying the following coupled \textit{linear} differential equations, known as the Bogoliubov-de Gennes (BdG) equations~\cite{2009PhRvL.102r0405Z,Zapata:2011ze}:
\begin{align}\label{BDG}
     \omega_j\begin{bmatrix}
    u_j\\
    v_j
          \end{bmatrix} =\begin{bmatrix}
          \hat{H}&-ge^{i\phi}\left|\Psi_0(x)\right|^2\\
          ge^{-i\phi}\left|\Psi_0(x)\right|^2&-\hat{H}^{*}
          \end{bmatrix}
          \begin{bmatrix}
    u_j\\
    v_j
                   \end{bmatrix},
\end{align}
where,
\begin{align}
    \hat{H}=-\frac{1}{2}\partial^2_x-\mu+V(x)+2g(x)\left|\Psi_0(x)\right|^2.
\end{align}
and $\phi$ denotes the phase of $\Psi_0^2$, taken to be uniform in view of the condition $\delta/r_{h}\ll 1$. Microscopic approaches to studying the superfluid ground state reveal that it is in fact a relative phase between different coherent occupations of pairing modes in the superfluid quantum ground state~\cite{Guo:2016ddz,fetter2003quantum}. Same holds for a Bardeen-Cooper-Schrieffer (BCS) superconductor~\cite{Bardeen:1957mv}, and permits a second-quantized description of Andreev reflections by treating the phase and charge/particle number as conjugate observables~\cite{Manikandan:2020kdx,PhysRevB.50.3139}. The more generic form of Eq.~\eqref{BDG} where $\omega\rightarrow i\partial_{t}$, and $u_{j}\rightarrow\bar{u}(x,t)$ and $v_{j}\rightarrow\bar{v}(x,t)$, assuming $\bar{u}(x,t)$ and $\bar{v}(x,t)$ are generic functions of $x$ and $t$ satisfy the following continuity equation:
\begin{align}\label{continuity}
    \partial_t\left(\rho_{u}-\rho_{v}\right)+\partial_{x}\left(J_{u}+J_v\right)=0,
\end{align}
where, the `currents' $J_{u/v}$ and `densities' $\rho_{u/v}$ are defined as:
\begin{eqnarray}
     \rho_{u}&=&\bar{u}(x,t)\bar{u}^*(x,t), ~~\rho_{v}=\bar{v}(x,t)\bar{v}^*(x,t),\nonumber\\
    J_{u}&=&\textrm{Im}\left[\bar{u}^*(x,t)\partial_{x}\bar{u}(x,t)\right],~~J_{v}=\textrm{Im}\left[\bar{v}^*(x,t)\partial_{x}\bar{v}(x,t)\right].\nonumber\\
\end{eqnarray}

An important physical insight about the nature of $u$ and $v$ modes is revealed by Eq.~\eqref{continuity}---while the excitation of a $u$-type mode introduces a positive change in the particle density of the condensate, that of a $v$-type mode induces a negative change. Hence, one might view the $v$-type modes as the bosonic analogue of holes. In fact, for this reason, we shall henceforth refer to the $v$-type modes as hole-like excitations/modes.
More arguments to substantiate this notion are also provided in~\cite{2009PhRvL.102r0405Z}. 

Also note that by performing integration by parts of Eq.~\eqref{continuity} with the boundary condition that currents vanish at infinity,  
we arrive at the condition, $\int (\rho_{u}-\rho_{v})dx=\text{constant}$, and the constant can be set to one as a normalization. Therefore the BdG equations also imply the Bogoliubov completeness relation for its components.   

When $\mu>0$, one can design a scenario in which the condensate is depleting via leaking to an ambient background of a coherent distribution of particles. Such a scenario was analysed in~\cite{2009PhRvL.102r0405Z} by Zapata and Sols, wherein it was also shown that the system exhibits a bosonic analogue of Andreev reflection.  To briefly review this finding here, we start by assuming the following simple form for the potential:
\begin{align}
    V(x)=Z\delta(x),\label{pot}
\end{align}
where $Z$ is the strength of the repulsive delta function at the interface. Further, assuming that the condensate is mostly confined to $x<0$, while the ambient distribution of particles is in the region $x>0$, the corresponding wave function can be approximated to:
\begin{align}
    \left|\Psi_0(x)\right|\approx \sqrt{n_c}\Theta(-x)+\sqrt{n_b}\Theta(x),\label{eqpsi}
\end{align}
where, $n_c$ and $n_b$ acquire the interpretation of particle density in the condensate and the ambient background regions, respectively and $\Theta(x)$ is the Heaviside-theta function. Following~\cite{2009PhRvL.102r0405Z}, we shall henceforth refer to the region of ambient background ($x>0$) as the normal side. Recall that on the condensate side, we have $n_cg(x)\rightarrow\mu$, while on the normal side, we have $n_bg(x)\rightarrow0$. With these inputs, one finds that a stationary wave function, of positive energy $\omega$, can be spanned by the set of vectors $\left\{\mathbf{e}_{u},\mathbf{e}_{v}\right\}$ on the normal side and $\left\{\mathbf{e}^{(+)}_{\omega},\mathbf{e}^{(-)}_{\omega}\right\}$ on the condensate side, where:
\begin{align}\label{sol1}
\mathbf{e}_{u}&=\begin{bmatrix}
    1\\
    0
\end{bmatrix},\qquad \mathbf{e}_{v}=\begin{bmatrix}
    0\\
    1
\end{bmatrix},\\
\mathbf{e}^{(\pm)}_{\omega}&=\frac{1}{\sqrt{e^{\pm2\theta_{\omega}}+1}}\begin{bmatrix}
    \pm e^{i\phi}e^{\pm\theta_{\omega}}\\
    1
\end{bmatrix}.
\end{align}
We have dropped the index $j$ for simplicity. We have introduced the parameter $\theta_{\omega}$ through the definition $\theta_{\omega}=\sinh^{-1}(\omega/\mu)$ and we have also retained the convenient $2\times 1$ matrix notation introduced in Eq.~\eqref{BDG}. 

\begin{widetext}
The particular stationary solution $\left(u_{\omega}(x),v_{\omega}(x)\right)$ that describes a stream of quasi-particles being incident on the condensate region from the normal side takes the following form:
\begin{align}
\mathbf{e}_{u}u_{\omega}(x)+\mathbf{e}_{v}v_{\omega}(x)=\begin{cases}
\mathbf{e}_u\left(e^{-ik^{+}_{\omega}x}+r_ne^{ik^{+}_{\omega}x}\right)+\mathbf{e}_{v}r_ae^{ik^{-}_{\omega}x}\quad&;\quad x>0\\
t_p \mathbf{e}^{(+)}_{\omega}e^{-ip^{-}_{\omega}x}+t_e\mathbf{e}^{(-)}_{\omega}e^{+p^{+}_{\omega}x}\quad&;\quad x<0
\end{cases}
\end{align}

\end{widetext}

where the different momenta are give as follows:
\begin{align}\label{k_plus_minus}
    k^{\pm}_{\omega}&=\sqrt{2\mu\left(1\pm\frac{\omega}{\mu}\right)},\\
    p^{\pm}_{\omega}&=\sqrt{2}\sqrt{\sqrt{\omega^{2}+\mu^{2}}\pm \mu}.
\end{align}The scattering amplitudes $r_n$, $t_p$ and $t_e$ correspond to that for normal reflection, transmission into the condensate and excitation of an evanescent mode, respectively~\cite{2009PhRvL.102r0405Z}. The amplitude of normal reflection $r_{n}$ will be nonzero for nonzero $Z$, the strength of the repulsive delta function at the interface. The situation we consider will have a non-zero $r_{n}$ by default, because one requires $Z\gg\sqrt{\mu}$ to ensure that the wavefunction in Eq.~\eqref{eqpsi} closely matches a stable solution of Eq.~\eqref{GP} and Eq.~\eqref{pot}~\cite{2009PhRvL.102r0405Z}. In addition, one finds that there is one more scattering amplitude, namely $r_a$, which is non-zero in general and, when $\mu>0$ and $|\omega|<\mu$ this amplitude corresponds to excitation of a propagating $v$-type (hole-like) mode on the normal side. Hence, the implication of $r_a\neq0$ is that there is a non-zero amplitude for the process of an incoming particle-like mode to get `absorbed' by the condensate, accompanied by the excitation of a propagating hole-like mode on the normal side---in close analogy to the well known Andreev reflection process in superconductor/normal-metal interfaces~\cite{osti_4071988}. In light of this, $r_a$ is referred to as the amplitude for Andreev reflection~\cite{2009PhRvL.102r0405Z}. An important observation concerning this amplitude, relevant to our discussion, is that the ratio $r_a/r_n$ has the following form:
\begin{align}\label{andreev_by_normal}
    \frac{r_a}{r_n}=e^{-i\phi}\left[e^{i\sigma(\omega,\mu,Z)}f(\omega,\mu,Z)\right],
\end{align}
where, $\sigma(\omega,\mu,Z)$ and $f(\omega,\mu,Z)$ are real valued functions of their arguments and independent of $\phi$.  
It also follows from the continuity equation [Eq.~\eqref{continuity}], and the completeness relation for $u$ and $v$ components that the ratio $|r_{a}/r_{n}|\leq 1$.   Eq.~\eqref{andreev_by_normal} also implies that Andreev reflection provides us a promising window to probe the relative phase added by a condensate, which in turn is related to the macroscopic quantum state of the condensate. In the superconducting case, it has been suggested that the phase added upon Andreev reflections will have measurable consequences in the current fluctuations observable in a superconductor-normal metal-superconductor junction~\cite{PhysRevB.50.3139}.

Experimental observation of Andreev reflections from a superfluid may face additional challenges, given that hole-like excitations are defined w.r.t an outgoing, coherent background of the superfluid. For example, various decoherence mechanisms in the superfluid background can lead to attenuation of a propagating hole-like mode, which is not accounted for in the discussions above. We may include this in the discussions by a phenomenological modification of the Andreev wavevector, $k_{\omega}^{-}\rightarrow k_{\omega}^{-}+i\kappa$, where $\kappa^{-1}$ is the length scale over which the superfluid background decohers.

Before we move on, we shall emphasize certain key points concerning our discussion so far.  Note that we have deliberately considered a leaky condensate. The reason for this is that we are interested in the study of quantum black holes, which are believed to be well described by leaky bosonic condensates of gravitons in certain models, as for instance in~\cite{Dvali:2012en}. For the bosonic system that we described in this   section, the leakage is guaranteed by the existence of hole-like propagating modes, with the corresponding momenta given by $k^{-}_{\omega}$, as in Eq.~\eqref{k_plus_minus}. Microscopically, the leakage may also be related to Andreev reflections from the interface, which allows to exchange modes between the normal fluid side and the superfluid; The time-reversal of a particle-like mode incident on the superfluid from the normal side is the Andreev reflection of a hole-like mode incident on the superfluid from the normal side, resulting in its mode-conversion into a particle-like mode. The depletion of a condensate via such a leakage can be identified with microscopic descriptions of black hole evaporation, as discussed in Refs.~\cite{Manikandan:2017zhw,Manikandan:2018urq,Manikandan:2020kdx,Jacobson:1996zs,tHooft:1996rdg}.

Next, we shall see how the above points guide us towards 
a fundamentally new kind of gravitational echo, when the near-horizon region of a quantum black hole is modelled as a superfluid quantum condensate. 

\section{Possibility of a new kind of gravitational echo\label{newkind}}
Equipped with important insights that we gained from our detour to a purely condensed matter scenario, we now return to our original problem of interest, viz. the black hole. Recall that we set off this article by introducing the Schwarzschild metric as the appropriate classical description of a spherically symmetric black hole spacetime. However, once we acknowledge that there is a finer description of a quantum black hole in terms of a graviton condensate, as for instance elaborated in~\cite{Dvali:2011aa,Dvali:2012en,Dvali:2012gb,Dvali:2012rt}, it is reasonable to picture the Schwarzschild metric as an effective, coarse-grained description. The nature of the corresponding fine-grained description is yet unknown owing, of course, to the unavailability of a consistent quantum theory of gravity. Despite this, however, it may be possible to realize a mean field description of the graviton-condensate picture of the black hole, say, in terms of an appropriate generalization of the Gross-Pitaevskii equation~\cite{Alfaro:2016iqs,Cunillera:2017kwe}. An interesting implication of this, which seems to have not been adequately appreciated previously, is that dynamics of the modes of perturbations of a black hole near the horizon must be described by an appropriate BdG-like equations, as opposed to, say, Eq.~\eqref{regge_wheeler}. Naturally, this must be understood as a suitable extrapolation of the fact that perturbation of a condensate is described by the BdG equations, as we have seen in the last   section. Let us see how one can proceed with such a proposal and, most importantly, make predictions with it.

Motivated by our analysis based on BdG equations in the condensed matter scenario, we expect that the modes of perturbations of the black hole, in the condensate picture, must be represented by a pair of functions $(u_{l,\omega}(x),v_{l,\omega}(x))$. In the following, we will suppress the $l$ index for simplicity. As in the case of bosonic condensate we considered in the previous   section, $u-$type excitations correspond to particle-like modes while $v-$type excitations correspond to hole-like modes. Note that the interpretation of a hole-like mode, in the black hole context, is that it describes a propagating negative particle density, with respect to the ambient background of gravitons furnished by Hawking radiation process. The coupled linear equation satisfied by $(u_{\omega}(x),v_{\omega}(x))$ is expected to depend on the details of near horizon scattering of gravitons. However, in order to understand the key qualitative features of this approach, we shall make some phenomenological assumptions. 

Note that, in the source free case, the near horizon limit of Eq.~\eqref{regge_wheeler} leads to the well known dispersion relation $k=\pm\omega$ for a propagating plane wave $e^{-i\omega t+ikx}$. In contrast, the existence of hole-like excitation, inferred from our condensed-matter based insights, manifest in the form of modification of the dispersion relation near $x=x_{\delta}+0^{+}$ to the effect of $k=\pm k^{\pm}_{\omega}$, where the index $\pm$ denotes $u$ and $v-$type modes, respectively. Therefore we expect that the near horizon region imparts the following boundary condition:
\begin{eqnarray}
 \begin{bmatrix}
        u_{\omega}(x)\\
        v_{\omega}(x)
    \end{bmatrix}&\propto&\mathbf{e}_{u}\left(e^{-ik_{\omega}^{+}x}+\mathcal{R}_{n}(\omega)e^{ik^{+}_{\omega}x}\right)\nonumber\\&+&\mathbf{e}_{v}\mathcal{R}_{a}(\omega)e^{ik^{-}_{\omega}x}\quad;\quad x\sim x_{\delta}. \label{echss}   
\end{eqnarray}
Similar to the superfluid/normal fluid example we discussed before, $\mathbf{e}_{u/v}$ can be understood as appropriate normal basis vectors in the region $x>x_{\delta}$. A modification similar to the first term in the above expression is well understood in some cases, for instance, in the context of testing the proposal that the horizon area is quantized~\cite{Bekenstein:2020mmb,Bekenstein:1995ju}, where the normal reflectively $\mathcal{R}_{n}(\omega)$ contain signatures of horizon area quantization, and encripts that in the gravitational echo signal~\cite{Cardoso:2019apo} (see Sec.~\ref{gechoes}). In comparison,  suggesting the second term in Eq.~\eqref{echss} proportional to $\mathcal{R}_{a}(\omega)$ resulting from mode conversions near the event horizon is the main contribution of this article. 

It is worth emphasizing that some of the existing proposals already point at such a modification, for instance, it has been pointed out that the near horizon region may facilitate mode conversions to resolve the Trans-Planckian reservoir problem at the event horizon~\cite{Jacobson:1996zs}. Superconducting and superfluid quantum information mirror analogies primarily motivated by the black hole quantum final state proposal in Ref.~\cite{Horowitz:2003he} also suggests such a modification to the boundary condition applied by the near-horizon region on test fields to resolve the black hole quantum information loss problem~\cite{Manikandan:2017zhw,Manikandan:2018urq,Manikandan:2020kdx}. 

We now proceed to discuss the observable consequences of the Andreev contribution---proportional to $\mathcal{R}_{a}(\omega)$ in Eq.~\eqref{echss}---on the gravitational echo signals. Although writing down the precise form of $\mathcal{R}_{a}(\omega)$ is beyond the scope of the present article, we make the following remarks based on the known physics of the condensed matter scenario discussed in Sec.~\ref{andreevbosons}.
Primarily, we expect that the echo produced in the condensate picture of a quantum black hole will have a particle-like and a hole-like component, possibly out-of-phase from each other, offering a new window to probe the quantum nature of black hole horizons through gravitational echo measurements. The ratio $\mathcal{R}_{a}/\mathcal{R}_{n}$ at the horizon is expected to have a structure similar to that of Eq.~\eqref{andreev_by_normal}, where $\mathcal{R}_{n}$ is required to be nonzero for a stable leaking superfluid condensate  of the kind discussed in Sec.~\ref{andreevbosons}~\cite{2009PhRvL.102r0405Z}. Moreover, since Andreev reflections are mode conversions facilitated by the ground state of a quantum superfluid, the corresponding contribution to the gravitational echo signal---if exists---will be an observable signature of quantum gravity at lower energies. We expect the quantum filtering effect of the potential barrier $V_{l}(r)$ resulting in the echo signal can also help resolve this low-energy signature of the near-horizon quantum state, and permit clever experimental schemes to detect the same.  

Confirming the presence of a hole-like component in the detected signal will require a waveform comprising of a minimum of four echoes. Additionally, the Andreev contribution to the echo may also be measurable as an enhancement in the particle-like component in every other echo (with a periodicity of two echos). The enhancement results from the Andreev reflected hole-like component getting reflected back from the potential barrier $V_{l}(r)$, subsequently Andreev reflecting at the horizon as a particle-like mode. As the phase added upon two Andreev reflections cancel each other, this will be observable as an enhancement of the normally-reflected component of the measured signal with a periodicity of two echoes. 

Finally, our analysis also suggests possible new experiments on the condensed matter side with exciting applications. Potential hills comparable to $V_{l}(r)$ may be engineered near superfluid/normal fluid interfaces, and superconductor/normal metal junctions. the dynamics of quantum fluctuations of a leaky condensate in such modified potentials can be used to probe echo-like signals from a superfluid (superconductor)/normal fluid (normal metal) boundary, and to investigate their quantum technology applications.

\section{Discussion\label{discuss}}
There are several interesting proposals that attempt to describe  black holes in terms of quantum condensates\cite{Dvali:2011aa,Dvali:2012en,Dvali:2012gb,Dvali:2012rt}. The idea that quantum state of a black hole may be effectively characterised by a simple many-body quantum wavefunction, such as the condensate ground state of a quantum superfluid, is also in harmony with the well known fact that black holes are characterized by very few parameters in their classical rendition (such as their mass, charge and angular momentum). Analyses along this line have also been previously employed to address some of the important issues in proposing a quantum theory of gravity---(1) the trans-Planckian reservoir problem at the event horizons~\cite{Jacobson:1996zs}, and (2) the black hole quantum information loss problem~\cite{Manikandan:2017zhw,Manikandan:2018urq,Manikandan:2020kdx}---both by assigning a mode-converting mirror property to the event horizon, facilitated by Andreev reflections, well known to condensed matter physicists. 

The present article looked at a possible observational implication for such proposals, within the framework of gravitational echoes. A minimal model to describe the black hole horizon as a condensate/normal fluid boundary reveals that echoes may be created by such spacetimes with both particle-like and hole-like components, going beyond the traditional gravitational echo scenario. Beyond the interest in black hole quantum physics, it is expected that the formalism may directly apply to a wide class of exotic compact objects. 

Before we conclude, we would like to point out some of the shortcomings of our proposal.
Primarily, note that although some independent arguments are presented as to how the black hole horizon may behave as a condensate facilitating mode conversions, the proposal still falls short of presenting an exact correspondence between the two fields. Therefore while the predictions we make offer a new paradigm to probe existing models,  they do not necessarily implicate that the models themselves accurately capture the quantum physics of black hole horizons. Secondly,
we use a mean-field approach developed in Ref.~\cite{2009PhRvL.102r0405Z} to describe Andreev reflections and extend it to the gravitational echo framework. While such an approach adequately captures the essential details of the scattering process, it is desirable that a microscopic description is also provided as Andreev reflections involve mode conversions between individual modes. We defer this analysis to a future work. 

Finally, it is likely that computation of the Andreev contribution to a gravitational echo may pose several challenges which the present article did not address. In addition, resolving the Andreev component from the other contributions to gravitational echoes may also present certain technical challenges.   Here we briefly summarize some of these issues that we think could be relevant.
\begin{itemize}
\item  As we have mentioned in Sec. \ref{newkind}, the Andreev contribution to the echo signal can be accounted for by considering a two-component vector-like waveform, with $u$ and $v$ components, as shown in Eq.~\eqref{echss}. The evolution of this 2-component object, in the first order perturbative approximation, is expected to be a second-order linear differential equation. Although in the present paper we have alluded to the expected form of this equation in the near-horizon limit, motivated by the analogy with a super-fluid/normal fluid interface, the explicit form of the equation should be dictated by, among possibly several other things, the details of scattering processes in the near-horizon limit. Such details are also expected to guide us to fix the correct near-horizon boundary conditions for the 2-component waveform. On the other hand, we expect that the boundary condition at $r\rightarrow\infty$ for the $u$-component to coincide with the standard outgoing boundary conditions and that of the $v$-component can be more challenging. A possible strategy is, once again, to gain insights from an appropriate condensed-matter analogue system. To this end, we expect, to some degree of approximation, that we can model the system by introducing an appropriate external potential on the normal side of the superfluid/normal-fluid interface.        
 
\item  Recall, that the Andreev reflected component is defined w.r.t an ambient coherent background of the leakage furnished by the early Hawking radiation. This poses two additional major challenges for computation and detection of the Andreev reflected component to the echo signal. First, various decoherence mechanisms in the background can adversely affect the propagation of the Andreev reflected component. A likely consequence is attenuation of the Andreev reflected component, which can be accounted phenomenologically as discussed in Sec.~\ref{andreevbosons}. 
The second point is the weakness of gravitational waves from Hawking processes itself. Although Andreev mechanism provide a modified boundary condition for the near horizon region based on proposed microscopic details of how a black hole may evaporate~\cite{Manikandan:2020kdx,Manikandan:2017zhw,Manikandan:2018urq,Jacobson:1996zs}, the contribution Hawking processes make to metric perturbations are still expected to be rather weak. This is so because the observable rate of Hawking processes would still be largely dictated by the principles of black hole thermodynamics~\cite{Page_2005}, and this rate is not expected to be challenged significantly by the microscopic details of near horizon processes.  

\end{itemize} 
In summary, while offering a new window to probe the quantum nature of black holes through gravitational echo measurements, resolving the Andreev contribution invites considerable further studies on the theoretical modeling of various gravitational echo scenarios, and their detection through feasible experiments.    
\section{Acknowledgements}
The work of SKM was supported by the Wallenberg Initiative on Networks and Quantum Information (WINQ). KR was supported by the Research Associateship of Indian Association for the Cultivation of Science (IACS), Kolkata, India. The authors acknowledge insightful comments from Andrew N. Jordan, Sayak Dutta, Sumantha Chakraborty, and Kabir Chakravarti. 
\appendix

\bibliography{New_BH_Echoes.bib}
\end{document}